\documentclass[conference]{IEEEtran}
\IEEEoverridecommandlockouts
\usepackage{cite}
\usepackage{amsmath,amssymb,amsfonts}
\usepackage{algorithmic}
\usepackage{multirow}
\usepackage{graphicx}
\usepackage{array}
\usepackage{enumitem}
\newcolumntype{L}[1]{>{\raggedright\arraybackslash}p{#1}}
\usepackage[caption=false, font=footnotesize]{subfig} 
\graphicspath{{Figures/}}
\usepackage{comment}
\usepackage{textcomp}
\usepackage{url}
\usepackage{xcolor}
\newcommand{\ie}{\textit{i.e.,} }

\def\BibTeX{{\rm B\kern-.05em{\sc i\kern-.025em b}\kern-.08em
    T\kern-.1667em\lower.7ex\hbox{E}\kern-.125emX}}
\begin{document}

\title{Comprehensive Analysis of Cellular Uplink Performance in a Dense Stadium Deployment}

\author{
\IEEEauthorblockN{
S.M. Haider Ali Shuvo\IEEEauthorrefmark{1}\IEEEauthorrefmark{3},
Hardani Ismu Nabil\IEEEauthorrefmark{2},
Joshua Roy Palathinkal\IEEEauthorrefmark{1}, \\
Muhammad I. Rochman\IEEEauthorrefmark{1},
and Monisha Ghosh\IEEEauthorrefmark{1}}
\vspace{3pt}
\IEEEauthorblockA{
\IEEEauthorrefmark{1}University of Notre Dame, USA,
\IEEEauthorrefmark{2}Sebelas Maret University, Indonesia.\\ 
\IEEEauthorrefmark{3}Corresponding Author Email: sshuvo@nd.edu}
}

\maketitle

\begin{abstract}
Uplink performance remains a critical limitation in modern 5G networks, where UEs have to balance limited transmission power against propagation challenges. We conducted extensive measurements in the University of Notre Dame's football stadium, which has a seating capacity of 80,000 spectators, evaluating network behavior under both unloaded (pregame) and severely congested (game day) conditions, with a focus on uplink performance. Analyzing PHY-layer metrics captured via the Rohde \& Schwarz QualiPoc, we show that high-frequency TDD bands in the uplink are severely bottlenecked in both the spectral and temporal domains. Despite transmitting near maximum 3GPP power limits, propagation loss inherent to high-frequency bands restricts UEs to low MCS indices and low PRB allocations, even in unloaded networks. This inability to achieve wideband allocation is further compounded by the significantly smaller number of uplink slots compared to downlink slots in TDD frames. Consequently, we observe a severe disparity between uplink and downlink: while high-frequency TDD bands carry the majority of downlink throughput, the network relies heavily on lower-frequency FDD bands for uplink. Additional measurements under favorable propagation conditions around a Verizon COW deployment located in the stadium parking lot also show that this limitation is not solely propagation-driven; rather, the duplexing scheme itself also plays a significant role. Even when TDD bands achieve higher or comparable MCS, FDD bands have a performance edge in the uplink due to the restrictive, downlink-heavy TDD architecture. These findings emphasize the indispensable role of low-frequency FDD spectrum in sustaining uplink capacity, providing insights that will help guide the design of next-generation wireless networks.
\end{abstract}


\section{Introduction} \label{sec:introduction}
Modern 5G cellular networks provide substantial improvements in downlink performance, but their uplink capacity remains a significant limitation~\cite{xu2020understanding,rochman2025comprehensive}. In the US, most newly deployed 5G bands use time division duplexing (TDD)~\cite{dahlman20205g}. In these bands, base stations benefit from high transmit power and larger downlink slot allocations. Conversely, user equipment (UEs) are restricted by limited transmit power and a very small fraction of uplink slots, which limits uplink capacity. At the same time, uplink demand is rapidly increasing due to applications such as live streaming, content creation, and video conferencing~\cite{ericsson_mobility_events}. Consequently, commercial 5G cellular networks often fail to meet user uplink expectations, especially during periods of high congestion~\cite{khan2025mature,ghoshal2022indepth,chmieliauskas2025evaluation}.

\begin{figure}[t]
  \centering
  \subfloat[Target sections \label{fig:target_sections}]{%
    \includegraphics[width=0.48\linewidth]{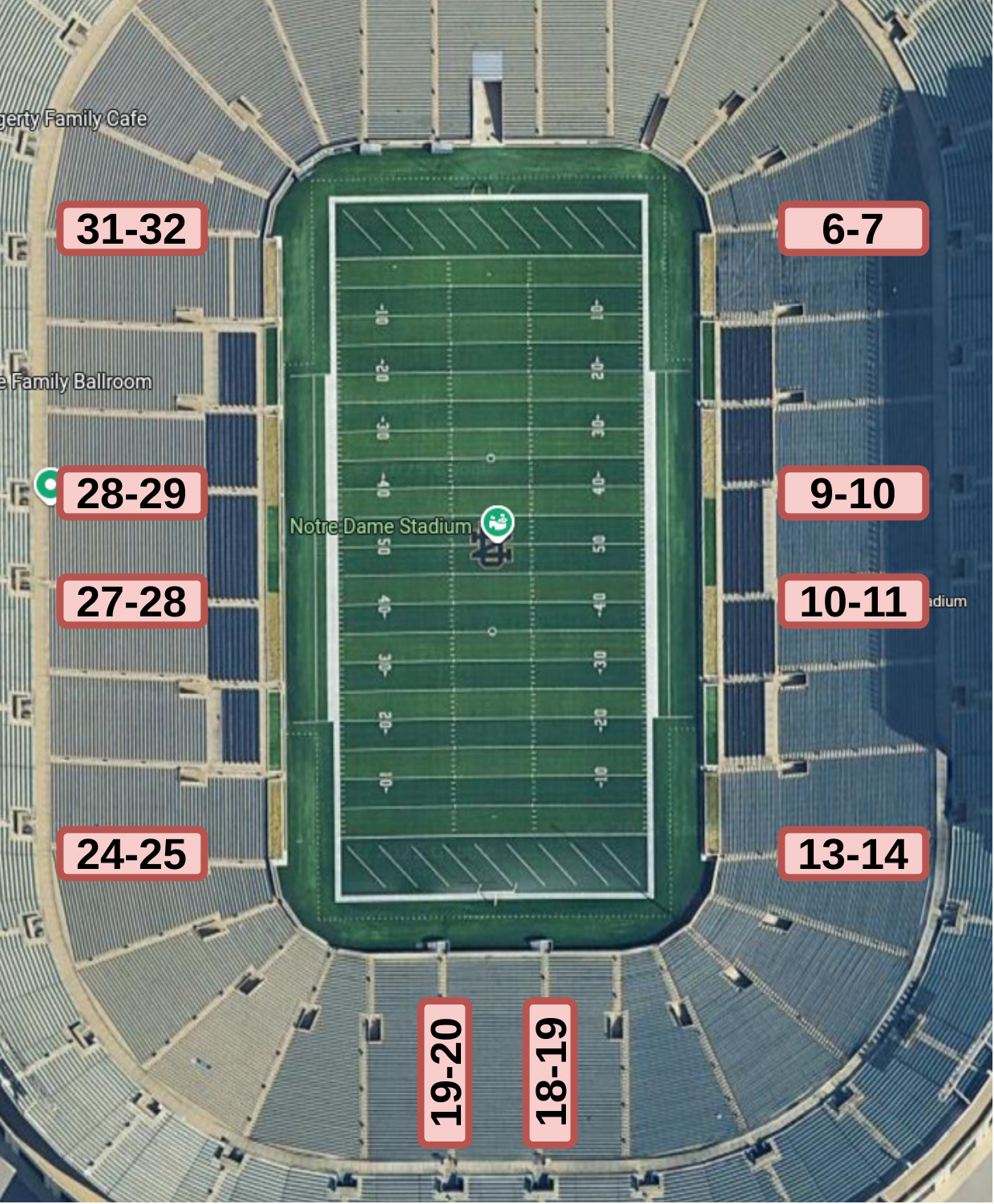}
  }
  \hfill
  \subfloat[Spots per section \label{fig:location_per_section}]{%
    \includegraphics[width=0.48\linewidth]{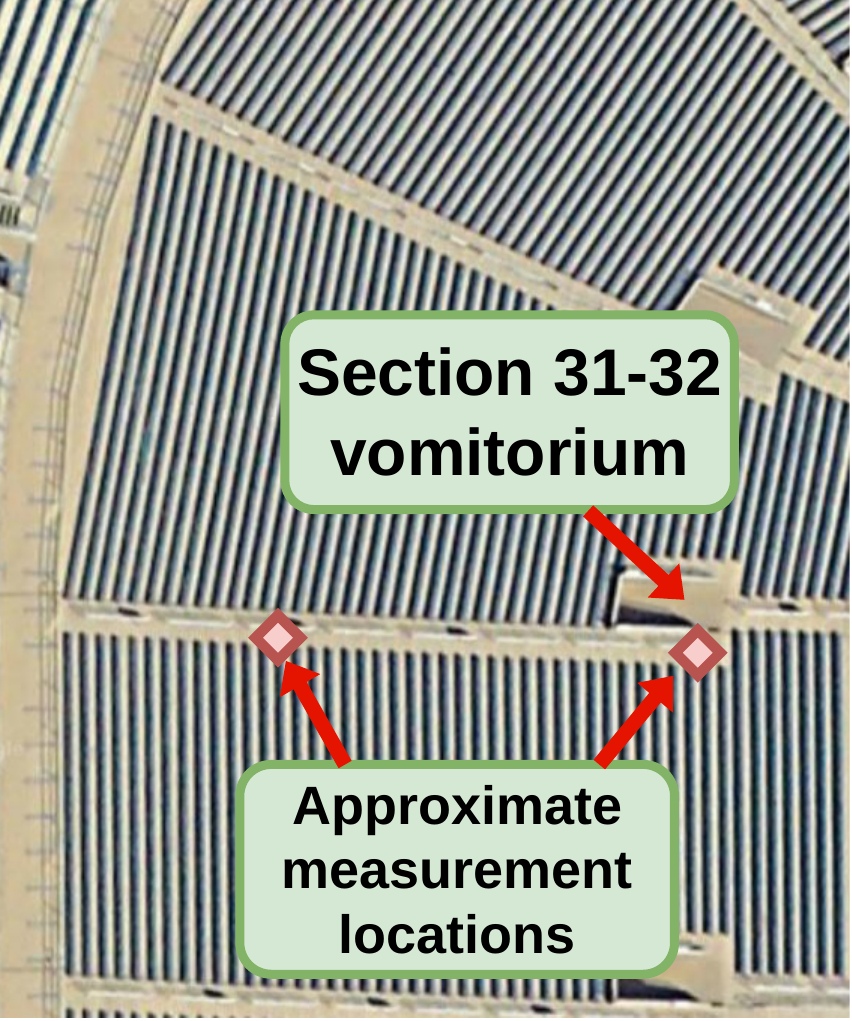}
  }
  
  \caption{Measurement spots at the stadium seating bowl.}
  \label{fig:measurement_target}
  \vspace{-1em}
\end{figure}

\begin{figure}[t]
    \centering
    \includegraphics[width=0.9\linewidth]{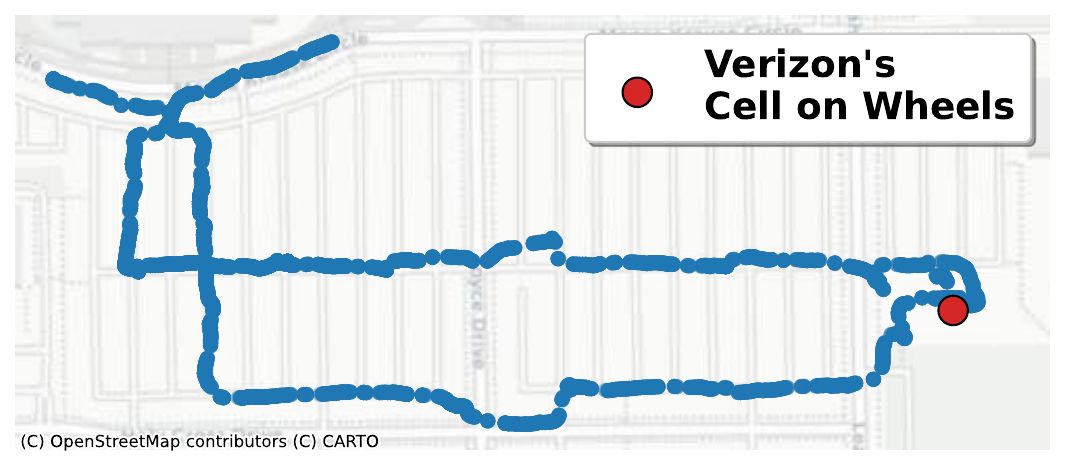}
    \caption{Measurement spots surrounding Verizon's Cell on Wheels (COW) located in stadium parking lot.}
    \label{fig:COW}
  \vspace{-1.5em}
\end{figure}

Modern stadiums, in particular, present a major example of ``uplink crunch'' for Mobile Network Operators (MNOs), with user densities often exceeding 50,000 active users per square kilometer~\cite{ookla_superbowl}. During games, many users engage in activities such as sharing videos, uploading content, video chatting, and live streaming, which severely stress the uplink~\cite{nokia_mass_event_2020}. Motivated by this challenge, we conducted extensive measurements at the University of Notre Dame football stadium during game time and in an empty stadium during the pregame period to better understand network performance in such scenarios. In addition, we collected and analyzed data around a Cell on Wheels (COW) deployed by an MNO in a parking lot outside the stadium, an approach operators increasingly use to handle high traffic demand~\cite{shakhatreh2021cell}. 
Through the analysis of these measurements, we make the following contributions:
\begin{itemize}[nosep, leftmargin=*]
    \item We present comprehensive real-world measurements comparing 5G Uplink performance in the congested stadium bowl and around a COW deployment in the parking lot.
    \item We show that uplink performance in high-frequency 5G TDD bands are severely limited by a combination of poor signal quality, reflected in low modulation and coding scheme (MCS) values, spectral constraints in which UE power limits large physical resource block (PRB) allocations, and a fundamental temporal bottleneck arising from the lack of available uplink slots in the frame structure. As a result, a larger fraction of uplink traffic is carried over the frequency division duplexing (FDD) anchor in commercial non-standalone (NSA) deployments. \looseness=-1
    \item Using measurements collected around the COW, where high-frequency TDD carriers achieve high MCS, we show that even when signal propagation is not a constraint, the uplink still relies heavily on the FDD carrier, indicating that slot allocation disparity is a major bottleneck.
\end{itemize}
We believe these findings provide insights for improving 5G network design and optimizing spectrum management toward better cellular uplink performance. In addition, by analyzing how the interplay of  frequency-dependent propagation and duplexing schemes impacts UE transmit power and performance, this work supports the goals of green communication to achieve more energy-efficient network configurations.

\begin{table}[t]
    \centering
    \caption{Captured LTE and NR parameters from QualiPoc.}
    \label{tab:qp_params}
    \begin{tabular}{|L{5em}|L{23em}|}
        \hline
        \textbf{Parameter} & \textbf{Description} \\
        \hline
        \hline
        Latitude, Longitude & UE’s geographic coordinates calculated by the Android API \\
        \hline
        PCI & Physical Cell Identifier \\
        \hline
        RSRP/\newline RSRQ & Signal strength values. For NR, RSRP/RSRQ indicates measurements from the 5G Synchronization Signal (SS) block. \\
        \hline
        DL/UL ARFCN & Downlink/Uplink Absolute Radio Frequency Channel Number, \ie center frequency. \\
        \hline
        PDSCH/\newline PUSCH Throughput & Throughput values recorded on the Physical Downlink Shared Channel and Physical Uplink Shared Channel, \ie PHY-layer downlink and uplink throughput. \\
        \hline
        DL/UL MCS & Modulation and Coding Scheme utilized in DL and UL channels.  \\
        \hline
        DL/UL RB & Number of allocated Resource Block per Subframe (LTE) or Slot (NR) in DL and UL.  \\
        \hline
        PUSCH Tx power & Uplink Transmit power used by the UE. \\
        \hline
    \end{tabular}
    \vspace{-1.5em}
\end{table}

\section{Data Collection Methodology}

Extensive measurements were conducted in the seating bowl of Notre Dame Stadium, which accommodates approximately 80,000 spectators during each game. Data were collected during two football games on Sep. 13 \& 20, 2025, denoted as \textbf{``Game Day''}, and, as a baseline, before the season started on Aug. 26 \& 28, 2025, denoted as \textbf{``Pregame''}. The data collection was performed using commercial smartphones equipped with Rohde \& Schwarz QualiPoc \cite{qualipoc}, a professional network measurement tool that logs low-level protocol events and key performance indicators (KPIs) by decoding Qualcomm modem signaling messages. A summary of the metrics captured by QualiPoc is provided in Table \ref{tab:qp_params}. 

Across all measurement days, we adhered to a standardized data collection protocol at each location: (i)~initialize a QualiPoc recording while stationary, (ii)~execute two consecutive Ookla speedtests, and (iii)~terminate the recording before relocating. Although Ookla Speedtest captures download, upload, and latency metrics, we omit the latency analysis in this paper due to space constraints. During the two game days, crowd-induced movement restrictions limited measurements to the stairways between seating sections. Consequently, we designated ten target sections and assigned two specific measurement spots per section: one near the vomitorium and another at the upper row. Fig.~\ref{fig:target_sections} illustrates the designated stairways, while Fig.~\ref{fig:location_per_section} shows the measurement spots within a representative section. In contrast, during the pregame phase, the absence of crowd restrictions allowed us to increase the number of spots while keeping them uniformly spaced. Across each measurement day, we recorded approximately 20 to 25 minutes of data per carrier. \looseness=-1

To accommodate high traffic demand during game time, Verizon deployed a COW in the stadium parking lot. In addition to the bowl measurements, we collected data in the accessible areas surrounding the COW during a game on Oct. 4, 2025. The COW features a three-sector deployment operating with a b66 primary carrier and a n77 secondary carrier. Due to site accessibility, our campaign was limited to two of these sectors, corresponding to Physical Cell Identifiers (PCI) 345 and 347 for the LTE and NR bands. The measurement footprint and the COW location are shown in Fig.~\ref{fig:COW}.

Our measurement setup consisted of two Samsung Galaxy S22+ smartphones and one Samsung Galaxy S24+ smartphone operating on the AT\&T, T-Mobile, and Verizon networks. All devices used premium, unthrottled data plans to capture peak throughput capabilities. We distributed the SIM card assignments across the devices on different measurement days to eliminate hardware bias. Due to UE hardware limitations, our setup did not support uplink carrier aggregation (CA) between two NR bands. Instead, uplink CA was limited to aggregation between the LTE anchor and a single 5G NR band in NSA mode. In SA networks, such as T-Mobile, the uplink relied entirely on single-carrier allocation without aggregation. 

\begin{table}[t]
    \centering
    \caption{Summary of LTE and 5G NR Carrier Configurations}
    \label{tab:carrier_config}
    \resizebox{\columnwidth}{!}{%
    \begin{tabular}{|c|c|c|c|c|c|c|c|c|}
        \hline
        \multirow{2}{*}{\textbf{Operator}} & \multirow{2}{*}{\textbf{Band}} & \multirow{2}{*}{\textbf{Duplex}} & \multicolumn{2}{c|}{\textbf{Freq. (MHz)}} & \textbf{BW} & \multicolumn{2}{c|}{\textbf{\# PCI}} & \textbf{$\Delta$PL} \\ \cline{4-5}\cline{7-8}
        & & & \textbf{UL} & \textbf{DL} & \textbf{(MHz)} & \textbf{PG\(^\dagger\)} & \textbf{GD\(^\dagger\)} & \textbf{(dB)} \\
        \hline
        \hline
        
        \multicolumn{9}{|c|}{\textbf{NR Bands}} \\
        \hline
        
        \multirow{1}{*}{Verizon} & n77 & TDD & \multicolumn{2}{c|}{3700, 3800} & 100 & 4 & 10 & {14.8} \\  
        \hline
        
        \multirow{3}{*}{T-Mobile} 
        & n71 & FDD & 677 & 631 & 20 & 0 & 1 & 0.0 \\ \cline{2-9}
        & n25 & FDD & 1894 & 1974 & 20 & 2 & 23 & 9.0 \\
        \cline{2-9}
        & n41 & TDD & \multicolumn{2}{c|}{2500, 2600} & 100, 90 & 17 & 37 & {11.3} \\ 
        \hline
        
        \multirow{1}{*}{AT\&T} & n5 & FDD & 828 & 873 & 10 & 21 & 23 & 1.7 \\
        \hline
        \hline
        
        \multicolumn{9}{|c|}{\textbf{LTE Bands}} \\
        \hline
        
        \multirow{4}{*}{Verizon} 
        & b13 & FDD & 782 & 751 & 10 & 23 & 24 & 1.3 \\
         \cline{2-9}
        & b5 & FDD & 840 & 885 & 10 & 10 & 24 & 1.9 \\ \cline{2-9}
        & b66 & FDD & 1720 & 2120 & 20 & 23 & 29 & 8.1 \\ \cline{2-9}
        & b2 & FDD & 1882 & 1962 & 5 & 23 & 21 & 8.9 \\
        \hline
        
        \multirow{2}{*}{T-Mobile} 
        & b66 & FDD & 1747 & 2147 & 15 & 2 & 23 & 8.2 \\
        \cline{2-9}
        & b2 & FDD & 1875 & 1955 & 10 & 1 & 23 & 8.8 \\
        \hline
        
        \multirow{4}{*}{AT\&T} 
        & b12 & FDD & 709 & 739 & 10 & 28 & 20 & 0.40 \\ \cline{2-9}
        & b66 & FDD & 1735 & 2135 & 10 & 25 & 19 & 8.2 \\
        \cline{2-9}
        & b2 & FDD & 1860 & 1940 & 20 & 27 & 24 & 8.8 \\ \cline{2-9}
        & b30 & FDD & 2310 & 2355 & 10 & 24 & 24 & 10.7 \\ 
        \hline
        
    \end{tabular}
    }
    \textit{\(^\dagger\) \textbf{PG} and \textbf{GD} refer to pregame and game day, respectively. \textbf{$\Delta$PL} denotes the free space path loss of the corresponding band relative to n71 uplink (677~MHz), calculated as $20\log_{10}(f_{UL}/677)$.}
    \vspace{-1.5em}
\end{table}

\begin{figure*}[htbp]
    \centering
    \subfloat[Uplink Transmit Power]{\includegraphics[width=0.48\linewidth, height=6.5cm]{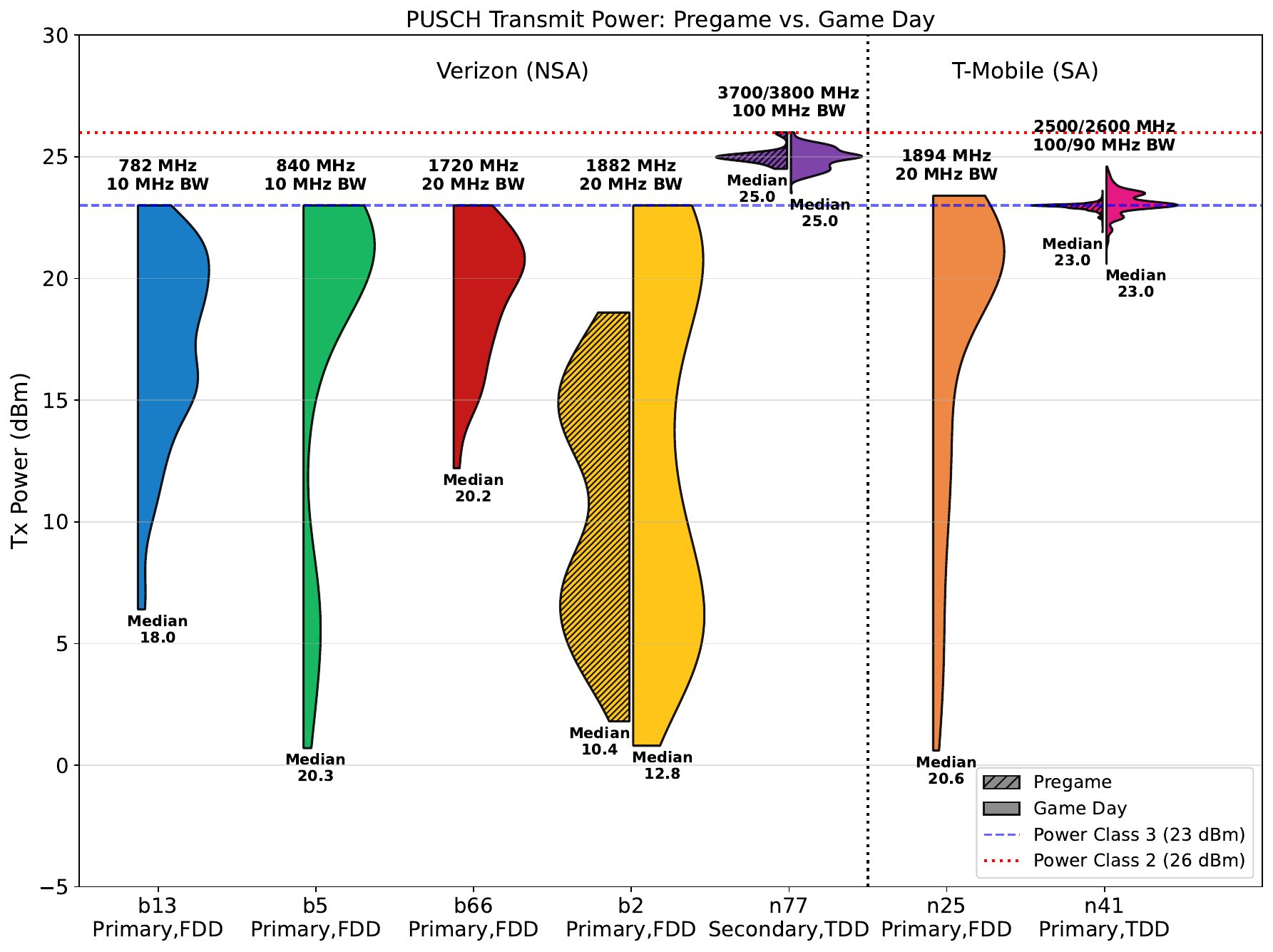}
    \label{fig:F1}}
    \hfil
    \subfloat[Uplink MCS]{\includegraphics[width=0.48\linewidth, height=6.5cm]{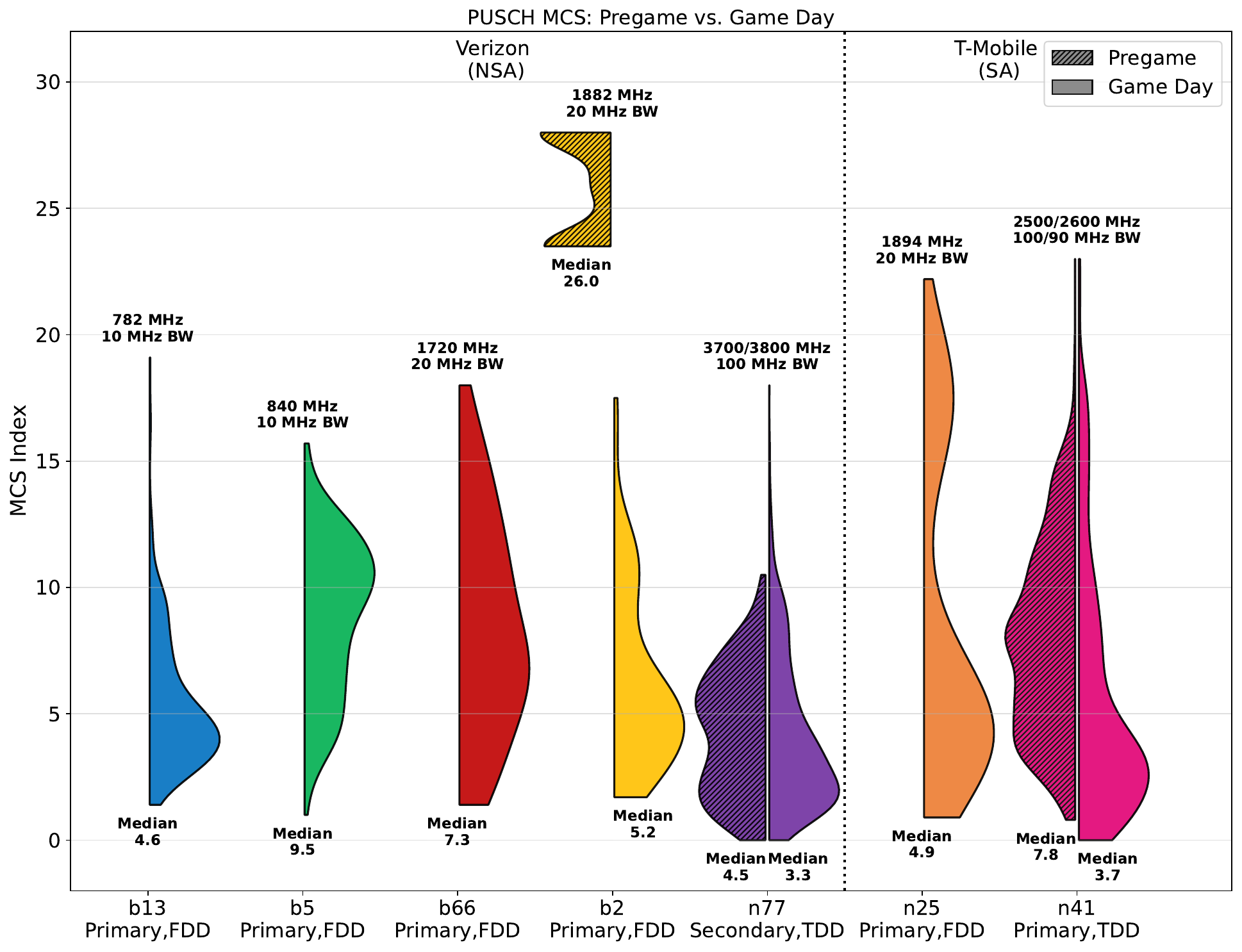}
    \label{fig:F2}}
    \caption{Uplink transmit power and MCS distributions across different bands under unloaded (pregame) and loaded (game day) conditions. Note the heavy concentration of transmit power near the 3GPP limit for high-frequency TDD bands (Verizon n77, T-Mobile n41), despite which the Uplink MCS remains severely degraded. Bands displaying only a right-sided violin indicate that the band that was only detected during game day conditions, with no corresponding data points during the pregame measurements.}
    \label{fig:tx_comparison}
    \vspace{-1.5em}
\end{figure*}

\section{Results}
The data collected during the two game days and the pregame period are analyzed to evaluate uplink network performance and quantify the inherent uplink-downlink asymmetry. The findings are presented in the following five subsections. Unless otherwise stated, the `Game Day' label represents the combination of the data from both the first and second games, and all presented results correspond specifically to data collected during active speedtest sessions.
\vspace{-2pt}

\subsection{Deployment Summary}
Verizon and AT\&T maintained 5G NSA connections throughout all measurement periods, while T-Mobile transitioned from exclusively Standalone (SA) mode during pregame to a roughly equal distribution of SA and NSA operation on game day. Only minimal instances of LTE-only operation were recorded for the operators. Table~\ref{tab:carrier_config} details the complete list of deployed bands active during game day, where the prefixes ``n'' and ``b'' indicate 5G and 4G bands, respectively. During the pregame, we observed limited spectrum utilization: Verizon served only b2 and n77 to our UEs, and T-Mobile utilized only n41. This stark contrast in active bands between the two phases highlights the severe spectrum crunch induced by game day user density. AT\&T's deployment remained similar in both conditions and had no C-band coverage in bowl area. \looseness=-1

Table~\ref{tab:carrier_config} also reports the number of unique per-band PCIs for pregame and game day. We observe a high PCI density on some bands during the game day, in particular TMO n25, n41, b66, and b2. This strongly indicates the use of active Distributed Antenna Systems (DAS) and/or small cells to address extreme capacity demands during the game. On the other hand, we observe a comparatively low number of Verizon n77 PCIs, which indicates lower number of small cells and/or passive DAS deployment.

\subsection{Uplink Power Consumption and Channel Quality}\label{sec:power_quality}
We analyze uplink transmit (Tx) power and uplink MCS distributions under empty pregame and packed game day conditions, as illustrated in Fig.~\ref{fig:F1} and Fig.~\ref{fig:F2}. 
For brevity, we focus this analysis on the Verizon and T-Mobile MNOs. Across both time periods, we observe that the high-frequency, wider TDD 5G bands (Verizon n77 and T-Mobile n41) utilize significantly higher transmit power than the lower frequency, narrow FDD bands (b5, b13, b66, b2, and n25). Specifically, T-Mobile's n41 consistently operates near the 3rd Generation Partnership Project (3GPP) Power Class 3 limit of 23 dBm, while Verizon's n77  approaches the Power Class 2 limit of 26 dBm \cite{3gpp_ts38101_1}. However, despite this higher power consumption, the median attainable MCS at these higher frequency bands remains considerably lower than that of the lower-frequency bands during both pregame and game day measurements. This indicates that the severe propagation loss inherent to high-frequency bands remains limiting even outdoors, thereby restricting the uplink to low MCS near the maximum permissible transmit power limit. 

\subsection{Uplink-Downlink Asymmetry in High-Frequency Bands}\label{sec:prb}
To evaluate the disparity between uplink and downlink, we analyze a single representative PCI observed on game day. We compare its MCS and PRB allocation ratio (defined as the ratio of allocated PRBs to the total available PRBs in that band) distributions across representative low- and high-frequency bands for both Verizon (b5 and n77) and T-Mobile (n71 and n41).

\begin{figure}[t]
    \centering
    \includegraphics[width=0.9\linewidth]{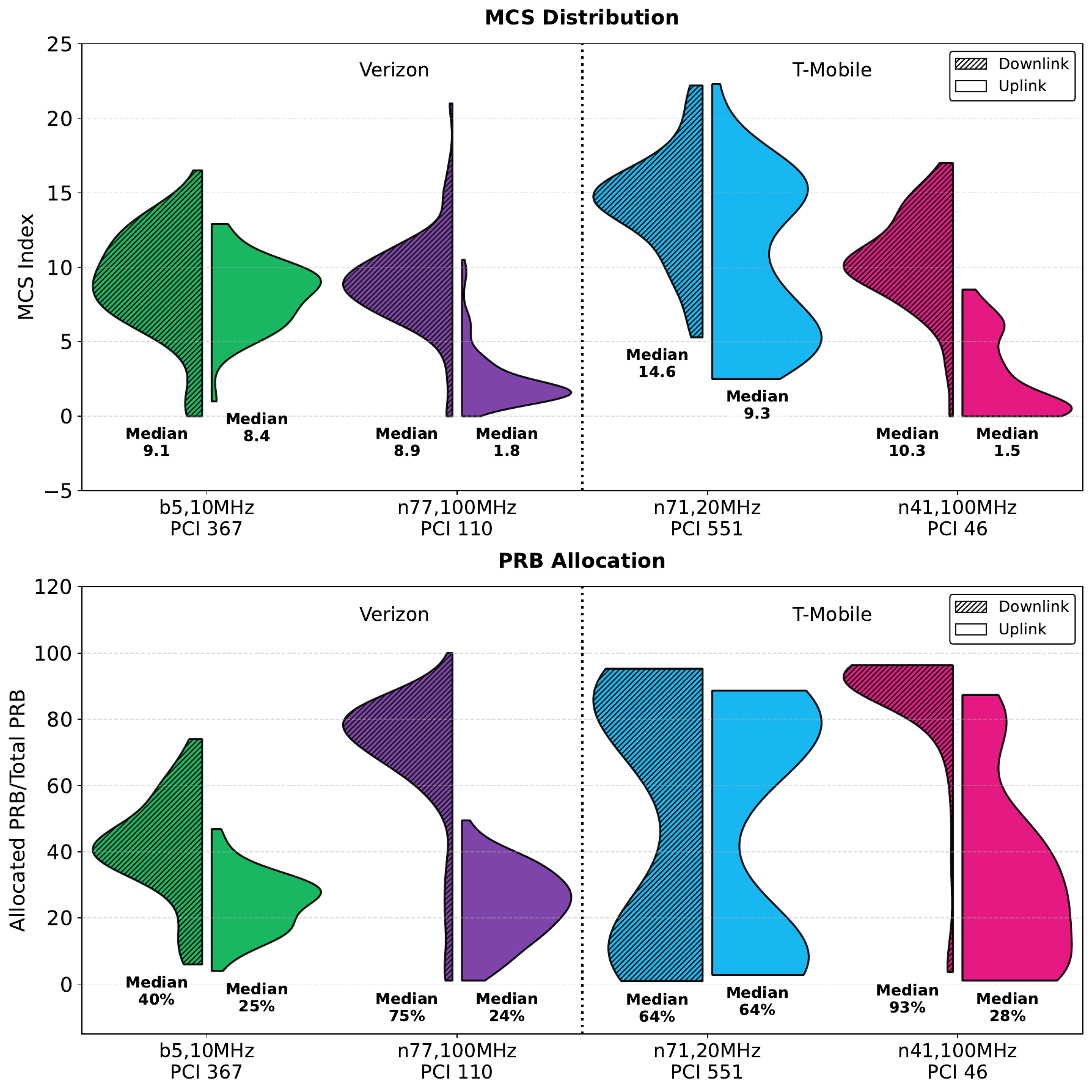}
    \caption{Comparison of downlink and uplink performance metrics on game day across identical PCIs for T-Mobile and Verizon's representative low- and high-frequency bands.}
    \label{fig:F3}
    \vspace{-14pt}
\end{figure}

Fig.~\ref{fig:F3} presents these distributions via split violin plots, where the left and right halves of each distribution denote the downlink and uplink, respectively. The upper subplot, detailing the MCS distribution, illustrates that while the base station successfully maintains a relatively high downlink MCS across all bands, the uplink MCS degrades severely in the higher-frequency bands (n77 and n41) under the same propagation conditions and the same PCI. 

Similarly, the lower subplot shows a significant reduction in the uplink PRB allocation ratio for the high-frequency bands. These observations highlight a fundamental asymmetry in the link budget: while the base stations can utilize their higher transmit power capability to counter high-frequency path loss in the downlink, the power-constrained UE fails to achieve a high MCS despite operating near its maximum permissible transmit power limit, as shown in \S\ref{sec:power_quality}. As a result, uplink performance in higher-frequency bands are severely bottlenecked compared to downlink, consistently failing to achieve high MCS or substantial resource block allocations.

\begin{figure}[t]
    \centering
    \includegraphics[width=0.75\linewidth]{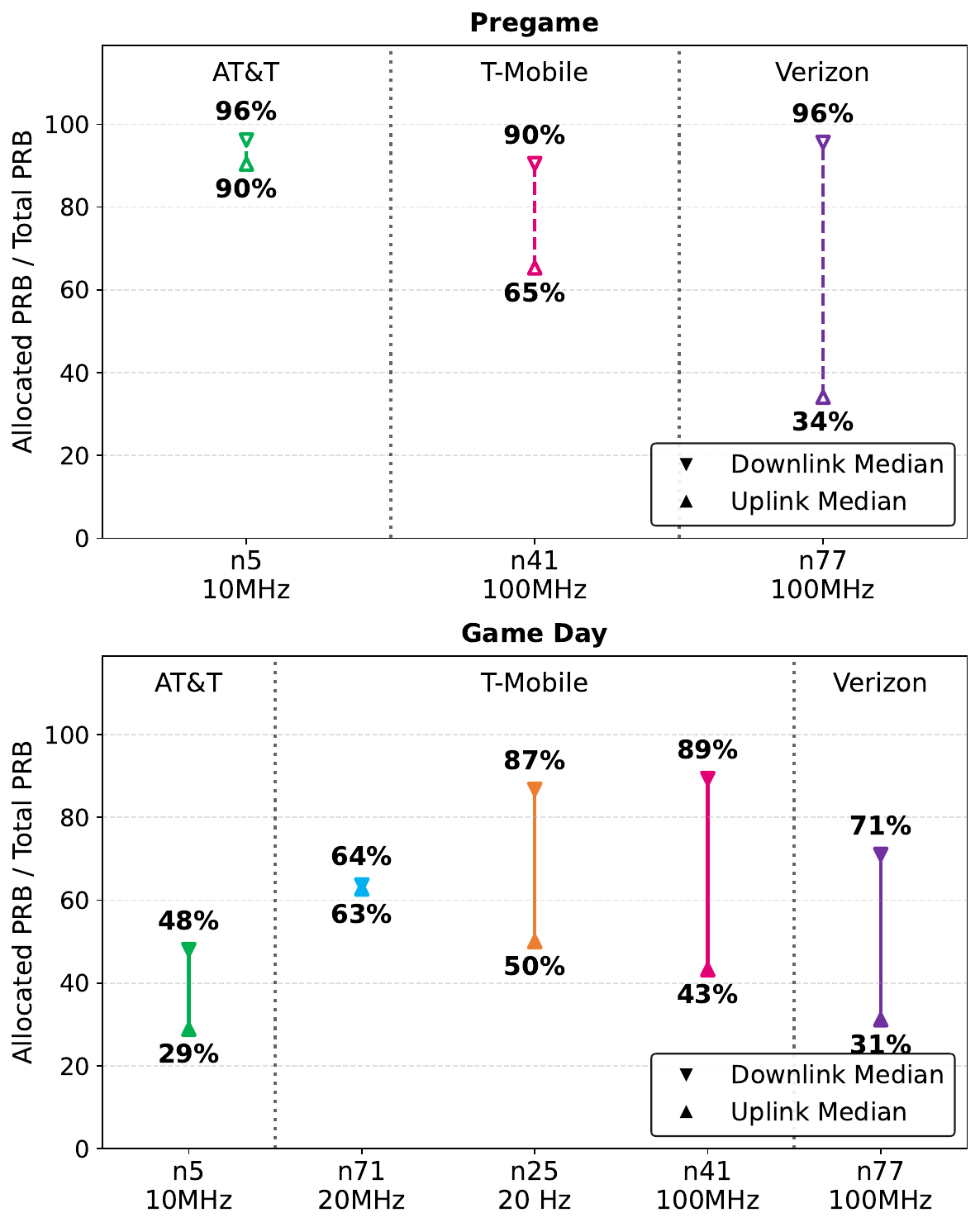}
    \caption{Median difference between downlink and uplink PRB allocation ratios across pregame (upper) and game day (lower) scenarios. The substantial gap in high-frequency bands between downlink and uplink even in the unloaded network highlights the inability of the UE to access the wideband high-frequency channel due to poor signal propagation and transmit power constraints.}
    \label{fig:F4}
\end{figure}

\begin{figure}[htbp]
    \centering
    \includegraphics[width=0.75\linewidth]{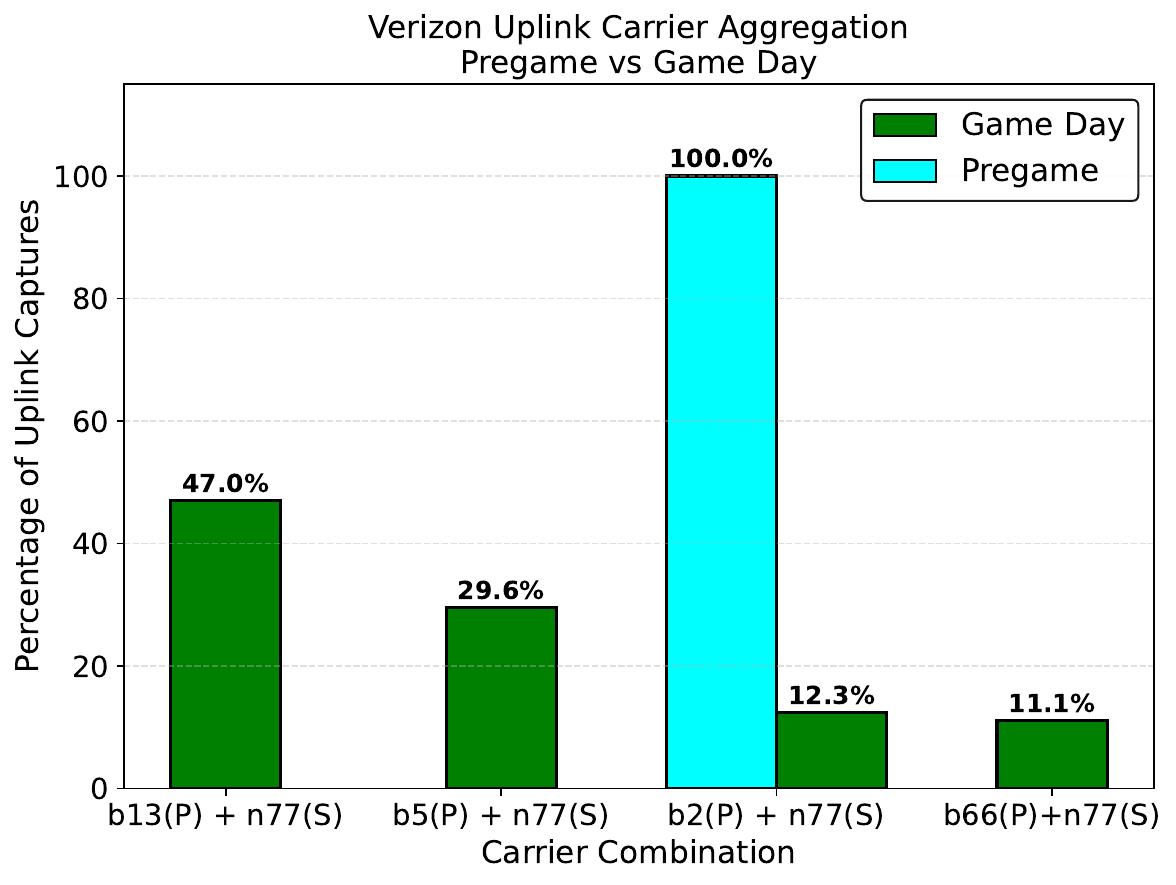}
    \caption{Histogram of Verizon uplink carrier combinations during pregame and game day. The distribution shows that during game day when network is congested and requires high capacity, the utilization of sub-1 GHz LTE anchors (b5, b13) becomes substantially more frequent.}
    \label{fig:F5}
    \vspace{-14pt}
\end{figure}

Fig.~\ref{fig:F4} illustrates the difference in median PRB allocation ratio between the downlink and uplink across the different 5G frequency bands of the three carriers during pregame and game day. We observe that this allocation gap is substantially larger for the high-frequency bands. Notably, this asymmetry persists even under unloaded pregame conditions, indicating that the uplink limitation is fundamentally power-limited rather than congestion-driven. As established in \S\ref{sec:power_quality}, the UE operates near its maximum 3GPP-allowable transmit power while still acheiving only a low MCS. Since these speedtest measurements were obtained in an empty stadium network, where radio resources are theoretically unconstrained, the reduced PRB allocation cannot be attributed to scheduler congestion. Instead, it indicates that the UE lacks sufficient total transmit power to sustain transmission across the full wide bandwidth of high-frequency carriers. To compensate for the severe propagation loss, the available uplink power is concentrated over a small subset of PRBs, thereby maintaining a viable uplink connection at low MCS in the high-frequency bands. \looseness=-1

\subsection{Prevalence of Lower-Frequency FDD Bands in Uplink Throughput}
To assess whether higher-frequency TDD bands contribute proportionally to the overall uplink throughput despite their higher transmit power and wider bandwidth, we first examine Verizon’s uplink performance in 5G NSA mode. As depicted in Fig~\ref{fig:F5}, during the congested game day scenario, the network relies heavily on sub-1~GHz carriers (b5, b13) as primary LTE anchors, accounting for 76.6\% of anchor usage, whereas during the pregame, only b2 (1882~MHz) served as the primary LTE anchor. This shift highlights the role of low-frequency FDD bands in sustaining uplink transmission under power-constrained conditions during periods of high user density.

\begin{figure}[t]
    \centering
    \includegraphics[width=0.80\linewidth]{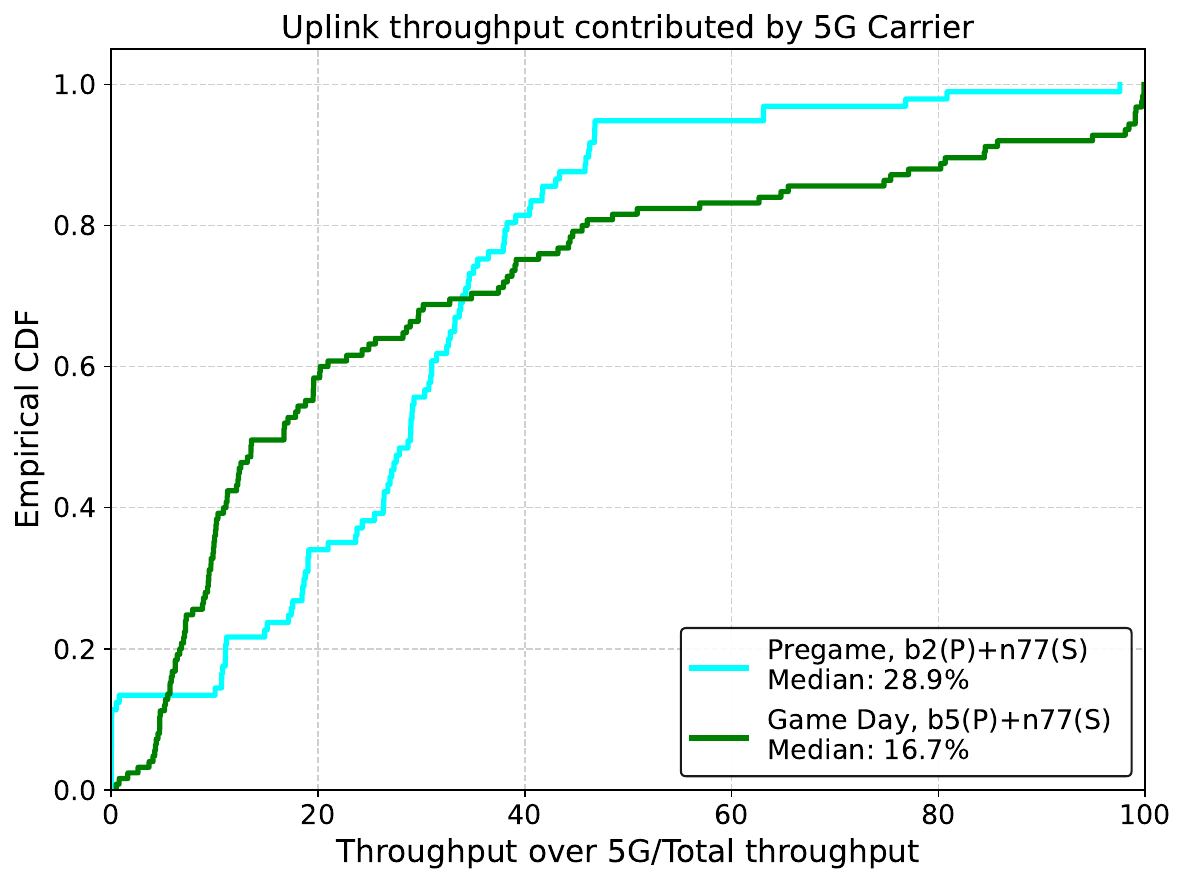}
    \caption{CDF plots illustrating the percentage of uplink throughput delivered by the Verizon 100 MHz, n77 carrier during pregame and game day. }
    \vspace{-14.5pt}
    \label{fig:F6}
\end{figure}

Furthermore, our measurements indicate that Verizon’s 5G NSA uplink performance depends heavily on the lower-frequency FDD bands. As shown in Fig.~\ref{fig:F6}, for an uplink CA configuration combining b5 (uplink 840~MHz, 10~MHz BW) and n77 (3700/3800~MHz, 100~MHz BW), the median 5G throughput contribution (5G throughput / total CA throughput) remains below 20\% during game day, even though n77 operates with a 6~dB higher median transmit power (Fig.~\ref{fig:F1}) and ten times the channel bandwidth. A similar trend is observed in the unloaded pregame stadium environment: when aggregating b2 (uplink 1882~MHz, 20 MHz BW) and n77, the median n77 throughput contribution remains below 30\%, despite utilizing approximately 14~dB higher median transmit power (Fig.~\ref{fig:F1}). As discussed in earlier sections, despite this higher transmit power, n77 in both game day and pregame exhibits low MCS values (Fig.~\ref{fig:F2}), suggesting that poor signal propagation at C-band limits its uplink capacity. \looseness=-1

\begin{figure}[htbp]
    \centering
    \includegraphics[width=0.9\linewidth]{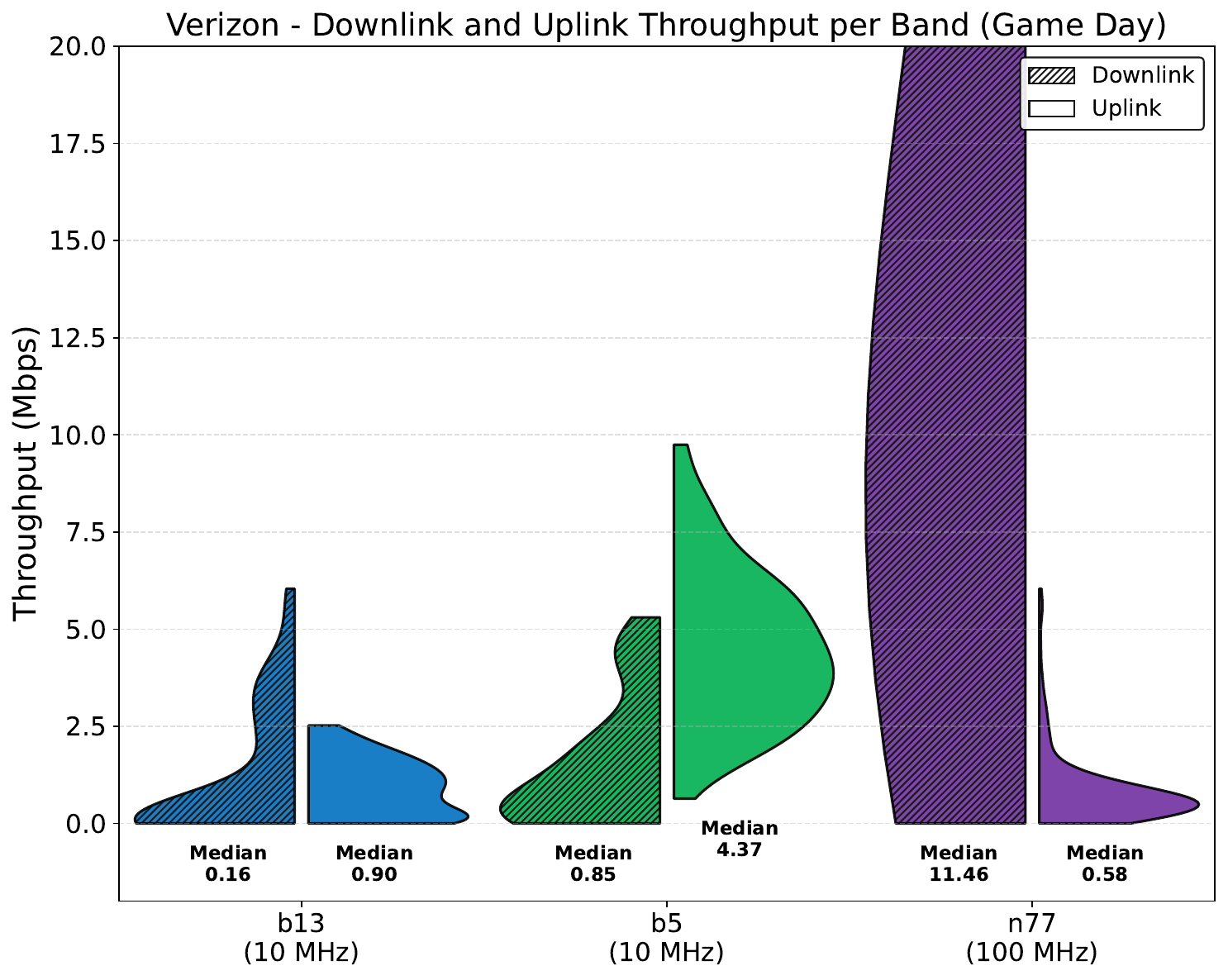}
    \caption{Downlink and uplink throughput distributions across b13, b5, and n77 bands during game day. The high-frequency TDD n77 carrier carries the vast majority of downlink traffic but fails to provide substantial uplink capacity, whereas the lower-frequency sub-1 GHz FDD bands (b5, b13) perform significantly better in uplink.}
    \label{fig:F7}
\end{figure}

A comparison of downlink and uplink throughput between the lower-frequency FDD bands (b5, b13) and the higher-frequency TDD band (n77) during game day further highlights the uplink-downlink asymmetry. As illustrated in Fig.~\ref{fig:F7}, while n77 achieves a high median downlink throughput of around 11.5~Mbps, its median uplink throughput drops to 0.5~Mbps. Conversely, the FDD bands exhibit much stronger uplink throughput relative to their downlink. Specifically, b5 exhibits median downlink and uplink throughputs of 0.9~Mbps and 4.4~Mbps, respectively, while b13 exhibits median values of 0.2~Mbps and 0.9~Mbps. Together, these results reveal a clear directional asymmetry in Verizon's deployment: the high-frequency TDD n77 band dominates downlink capacity, while uplink throughput depends heavily on the sub-1~GHz FDD bands (b5, b13) during game day.

\begin{figure}[htbp]
    \centering
    \includegraphics[width=0.80\linewidth]{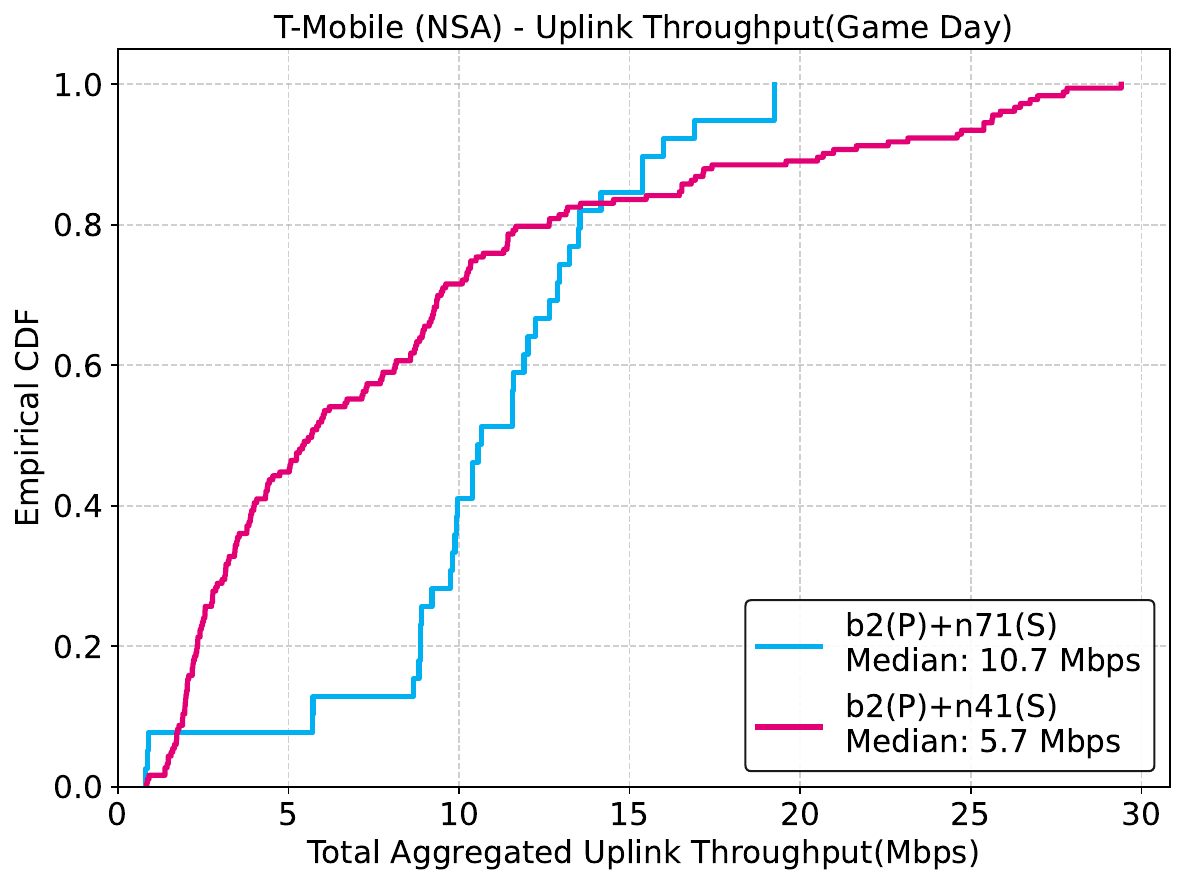}
    \caption{Total aggregated uplink throughput CDF during game day for T-Mobile NSA mode. Notably, the network achieves a higher median throughput when low-frequency n71 FDD band is aggregated as the secondary carrier alongside the b2 anchor, rather than the high-frequency n41 TDD band.
    }
    \vspace{-13.5pt}
    \label{fig:F9}
\end{figure}

Next, we evaluate T-Mobile’s uplink performance when operating in 5G NSA mode (a discussion of T-Mobile's 5G SA mode is provided in \S\ref{sec:fdd_adv}). Unlike Verizon, which relies solely one the high-frequency TDD n77 band for 5G stadium coverage, T-Mobile deploys both a low-frequency FDD 5G carrier, n71 (uplink 677~MHz, 20~MHz BW), and a high-frequency TDD 5G carrier, n41 (2500/2600~MHz, 100/90~MHz BW). This dual deployment allows a direct comparison of aggregated NSA uplink throughput across the two frequency tiers. As illustrated in Fig.~\ref{fig:F9}, although aggregating with n41 enables a higher peak uplink throughput, the carrier combination of b2 with n71 achieves a significantly higher median aggregated uplink throughput (10.7~Mbps) than b2 and n41 (5.7~Mbps).

Consistent with the uplink-downlink asymmetry observed in Verizon's deployment, T-Mobile's high-frequency carrier exhibits a similar performance gap. When examining per-band downlink and uplink performance, 
we observe that while the high-frequency n41 band delivers a median downlink throughput of 44~Mbps, its median uplink throughput drops sharply to 1.5~Mbps. Conversely, the lower-frequency bands exhibit a much more balanced profile: n71 records median downlink and uplink throughputs of 20~Mbps and 10~Mbps, respectively, while the LTE anchor b2 achieves 3~Mbps and 1~Mbps. Similar to Verizon's deployment, the T-Mobile data shows the same trend: the high-frequency TDD n41 band dominates downlink capacity, while the low-frequency FDD n71 carrier provides higher uplink throughput.

Taken together, the Verizon and T-Mobile results reveal a consistent operational pattern in 5G networks: high-frequency TDD carriers are fundamentally uplink-limited by three compounding constraints:
\begin{enumerate}[nosep, leftmargin=*]
    \item From a propagation perspective, the larger path loss at high frequencies degrades channel quality, yielding low uplink MCS despite operation near the 3GPP transmit power limit, as shown in \S\ref{sec:power_quality}.
    \item From a spectral perspective, UE transmit power constraints prevent effective utilization of the wideband high-frequency channels, resulting in substantial PRB allocation disparity relative to the downlink. While the base station utilizes a large transmit power budget to maintain high PRB allocation in the downlink, the power-constrained UE is unable to secure a comparable allocation in the uplink, even under unloaded conditions, as shown in \S\ref{sec:prb}.
    \item From a temporal perspective, downlink-heavy TDD slot configurations further restrict uplink transmission opportunities. Consequently, although high-frequency TDD carriers dominate downlink throughput, uplink performance in both T-Mobile and Verizon NSA deployments remains heavily dependent on low-frequency FDD bands.
\end{enumerate}

\begin{figure}[t]
    \centering
    \includegraphics[width=0.8\linewidth]{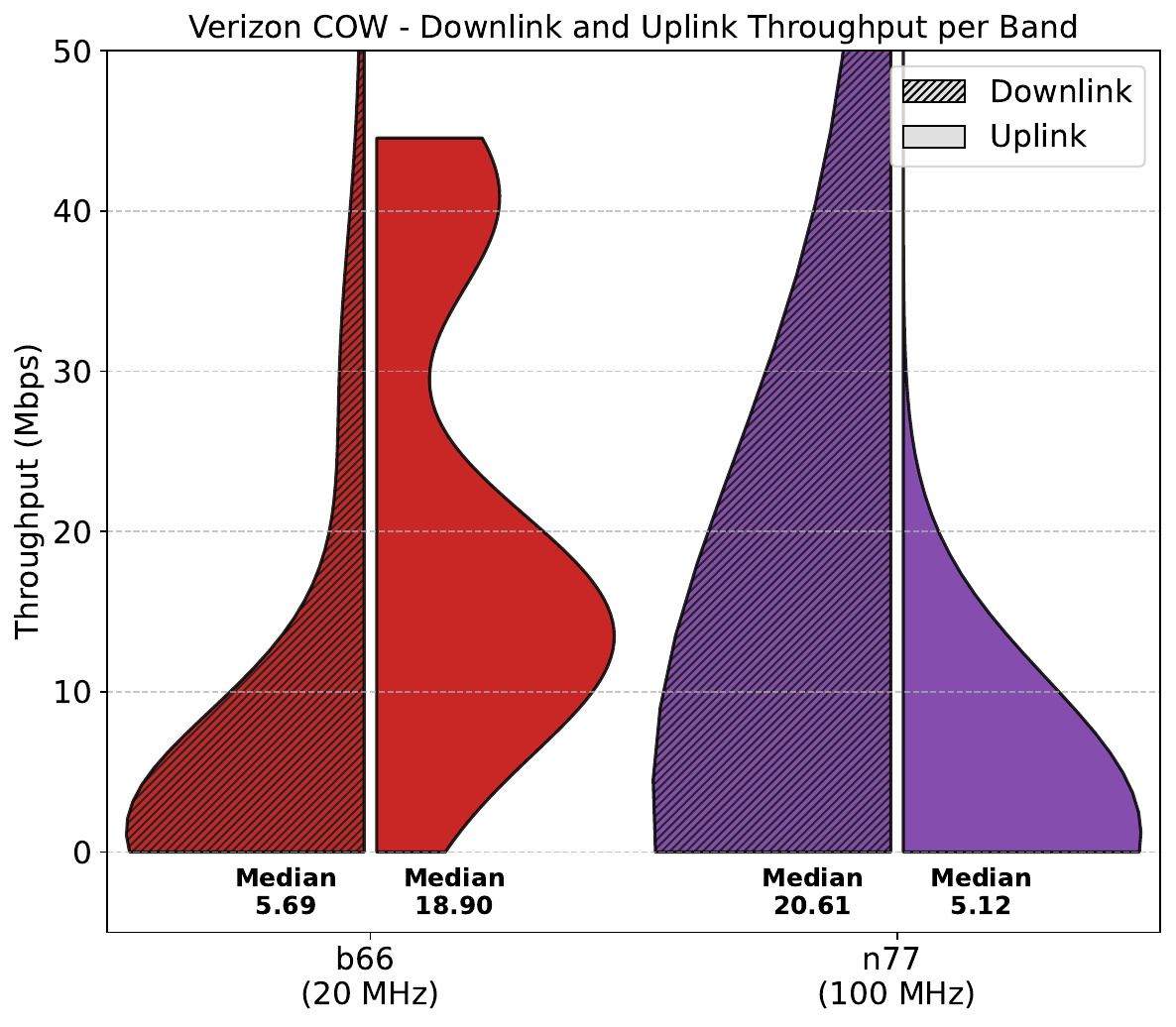}
    \caption{Downlink and uplink performance comparison on the Verizon COW utilizing a primary b66 anchor and secondary n77 band. Despite conducive C-band propagation conditions (for band n77 the median  MCS: UL 20/DL 16; median UL Tx power: 19 dBm), it shows a disproportionate throughput distribution. While n77 carries the bulk of the downlink load, the b66 anchor provides a median uplink throughput approximately four times higher. }
    \vspace{-14pt}
    \label{fig:F10}
\end{figure}

\subsection{Performance Advantage of FDD in Uplink}\label{sec:fdd_adv}
To assess whether high-frequency 5G TDD carriers can sustain uplink loads comparable to or greater than those carried by low-frequency FDD bands under favourable propagation conditions and high MCS, we analyze measurements collected around the COW deployment during the Oct. 4, 2025, game day for PCIs 345 and 347.

Given the close proximity of the measurement locations to the COW and the high likelihood of line-of-sight propagation, both the primary b66 band (uplink 1720~MHz, 20~MHz BW) and the secondary n77 band (3700/3800~MHz, 100~MHz BW) achieve high MCS values in both downlink and uplink while operating substantially below the 3GPP Power Class 2 limit of 26~dBm (the median uplink transmit power is 19~dBm for n77 and 17~dBm for b66). Specifically, the uplink median MCS is approximately 18 for b66 and 20 for n77. However, despite this favorable MCS on n77, Fig.~\ref{fig:F10} shows that the median n77 uplink throughput remains near 5~Mbps, which is roughly one-fourth of the uplink throughput carried by b66. This behavior reverses in the downlink, where n77 carries the vast majority of the load. The 5G contribution ratio further confirms this asymmetry: although n77 carries nearly all downlink traffic (median 97\%), its median uplink contribution remains below 10\%. Moreover, the median uplink PRB allocation is about 10\% lower than the downlink allocation, despite substantial UE transmit power headroom, indicating that uplink resource contention remains significant.

\begin{figure}[t]
    \centering
    \includegraphics[width=0.865\linewidth]{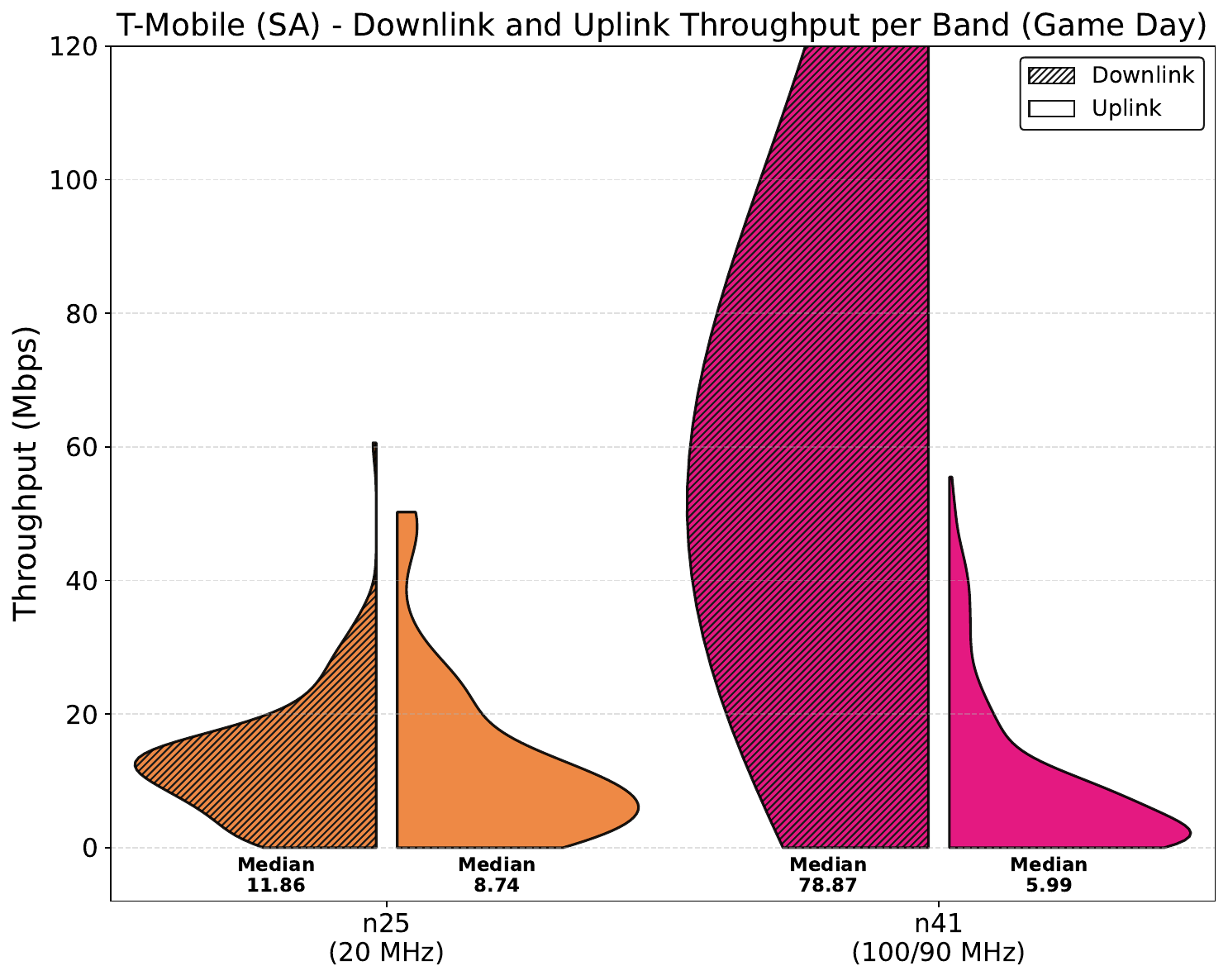}
    \caption{Downlink and uplink throughput distributions for T-Mobile's 5G SA mode, comparing the n25 (FDD) and n41 (TDD) carriers. Despite operating under comparable MCS conditions and the n41 band utilizing 3dB more uplink transmit power, the n25 FDD carrier maintains a distinct performance edge in the uplink, achieving a median throughput of approximately 9 Mbps compared to 6 Mbps for n41.}
    \vspace{-12pt}
    \label{fig:F12}
\end{figure}

The Verizon COW deployment indicates that even under excellent signal propagation and high MCS conditions---where the UE possesses significant transmit power headroom---the network remains heavily reliant on the primary FDD band for uplink capacity. The TDD carrier is firstly structurally limited by its downlink-heavy slot allocation, which reduces temporal uplink allocation. Furthermore, this temporal restriction induces a secondary spectral limitation under load: because the TDD frame provides a low number of uplink slots, user congestion is amplified in the uplink compared to the downlink during high-traffic events, as suggested by their median PRB allocation difference of $10\%$. Consequently, even with conducive propagation condition and Tx power not being a bottleneck, UE fails to achieve an uplink high enough PRB allocation on the wide TDD carrier comparable to its downlink, which combined with the inherently low number of uplink slots, forces the network to depend on the continuous FDD band for uplink capacity. \looseness=-1

A parallel analysis can be applied to T-Mobile’s 5G SA network. While T-Mobile does not deploy a COW, their SA architecture within the stadium bowl includes both a mid-frequency FDD band n25 (uplink-1894 MHz, 20 MHz BW) and a high-frequency TDD band n41 (2500/2600~MHz, 100/90~MHz BW). Without uplink CA, these carriers are allocated independently. As shown in Fig. \ref{fig:F2}, the n25 and n41 bands achieve comparable uplink MCS, with medians of approximately 5 and 4, respectively. In particular, the transmit power allocated to the n25 band is on average 3 dB lower than that of n41 (Fig \ref{fig:F1}). Examining the throughput per band (Fig. \ref{fig:F12}) reveals similar disparity like Verizon. Although n41 delivers far superior downlink performance---carrying roughly eight times the median throughput of n25---the n25 band has performance edge over n41 in the uplink, achieving a median of approximately 9 Mbps compared to n41's 6 Mbps. This occurs despite n41 utilizing twice the transmit power and both bands having comparable MCS. This reinforces the finding that high-frequency TDD bands are inherently less suitable for uplink traffic compared to FDD bands even when signal propagation and channel conditions are comparable, due to TDD bands asymmetric frame configurations that heavily favor downlink transmission.

\section{Conclusion}
In this paper, we presented a comprehensive measurement study of network performance in a high-density stadium environment, focusing on the fundamental performance contrast between the downlink and uplink. Our analysis reveals that uplink performance on high-frequency wide TDD bands is severely power constrained. Due to extreme propagation losses, UE consistently yields very low MCS and minimal PRB allocations compared to downlink, despite operating near the maximum permissible 3GPP transmit power limits. This leaves UE being unable to use much of the bandwidth even in situation of no network congestion like empty stadium.

Analyzing the NSA deployments of both Verizon and T-Mobile, stark asymmetry emerges in their downlink and uplink behavior. While wide, high-frequency TDD carriers bear the vast majority of the downlink load, the lower-frequency FDD bands are carrier of the bulk of the uplink traffic. Furthermore, we demonstrated that this uplink bottleneck is not strictly a product of poor propagation. By evaluating scenarios with excellent channel conditions (Verizon COW deployment) or with comparable channel conditions between FDD and TDD bands (T-Mobile SA deployment), we pointed the structural limitations of the TDD duplexing scheme. Even when high-frequency TDD bands achieved exceptionally high or comparable MCS values to their FDD counterparts, the FDD bands maintained a distinct performance edge in the uplink. This highlights that the asymmetric, downlink-heavy frame configuration of TDD bands fundamentally limits uplink capacity, regardless of signal quality.

Finally, these findings provide a critical insight regarding the impact of duplexing scheme on power constrained uplink. The TDD strategy relies on granting the UE intermittent, time-multiplexed access to a large bandwidth, with the uplink receiving a small fraction of the temporal slots compared to the heavily favored downlink in most deployments. However, this yield diminishing returns because as our measurements show the power-constrained UE cannot physically access such a wide channel under typical propagation conditions, leaving much of the band inaccessible even when the network is unloaded. In contrast, the FDD strategy allocates a smaller, dedicated chunk of bandwidth that is continuously available for uplink. This constant availability, combined with a narrower channel that is accessible with the UE's limited power budget, ensures the resource blocks can be better accessed both spectrally and temporally, making FDD architectures more advantageous and efficient for achieving uplink performance in real world scenarios.
\section*{Acknowledgments}
This work was supported by the National Science Foundation (NSF) under Grants ECCS-2434134 and CNS-2229387. The authors also gratefully acknowledge the support of a research gift from Cisco. We would also like to extend our gratitude to Mike Atkins for the technical support.

\bibliographystyle{IEEEtran}
\bibliography{references}

\end{document}